\documentclass[11pt]{article}

\usepackage{amssymb}

\begin{document}
\thispagestyle{empty}
\begin{center}
\vspace{2cm}
{\bf Supersymmetric hybrid inflation\\ 
\bf in the braneworld scenario}\\
\vspace{2cm}
H. Boutaleb-J\\
{\it Laboratoire de Physique Th\'eorique, Facult\'e des 
Sciences,\\
BP 1014, Agdal, Rabat, Morocco,}\\%[1.5em] 
A. Chafik\\
{\it Laboratoire de Physique Th\'eorique, Facult\'e des 
Sciences,\\
BP 1014, Agdal, Rabat, Morocco,}\\%[1.5em]
A. L. Marrakchi\\
{\it Laboratoire de Physique Th\'eorique, Facult\'e des 
Sciences,\\
BP 1014, Agdal, Rabat, Morocco,}\\
and \\
{\it D\'epartement de Physique, Facult\'e des Sciences,\\
BP 1796, Atlas, F\`es, Morocco}

\end{center}
\vfill

\vspace{1cm}
\begin{center}
{\Large\it To appear in Physics Letters B}
\end{center}

\newpage

\centerline{\bf Abstract}
\baselineskip=20pt
\bigskip

In this paper we reconsider the supersymmetric hybrid inflation in the context of the braneworld scenario . The observational bounds are satisfied with an inflationary energy scale $\mu\simeq 4\times 10^{-4}M_p$ , without any fine-tuning of the coupling parameter, provided that the five-dimensional Planck scale is $M_5\stackrel{<}{\sim} 2\times 10^{-3}M_p$ . We have also obtained an upper bound on the the brane tension .   \\
PACS number(s): 98.80.Cq

\newpage

Recently there has been a great deal of interest in conceiving our universe to be confined in a brane embeded in a higher dimensional space-time \cite{key1}. Such models are motivated by superstring theory solutions where matter fields (related to open string modes) live on the brane, while gravity (closed string modes) can propagate in the bulk \cite{key2}. In these scenarios extra dimensions need not be small and may even be infinite \cite{key3}. Another important consequence of these ideas is that the fundamental Planck scale $M_{4+d}$ in $(4+d)$ dimensions can be considerably smaller than the effective Planck scale $M_p = 1.2\times 10^{19}$ GeV in our four dimensional space-time.\\

It has also been noticed that in the context of extra dimensions and the brane world scenario the effective four-dimensional cosmology may deviate from the standard Big-Bang cosmology \cite{key4,key5}, which could lead to a different physics of the early universe . Thus, it is interesting to study the cosmological implications of these new ideas . Inflation is certainly the best framework for describing the physics of the early universe \cite{key6}, since it provides a consistent explanation of the standard cosmological puzzles like the horizon and flatness ones, the observed degree of isotropy in the cosmic microwave background radiation (cmbr) has also been succesfully explained in the framework of inflation. Furthermore, the Linde's hybrid inflation model \cite{key7} where more than one field are relevent has become the standard paradigm for inflation, in particular it is more useful for supersymmetric inflationary medels.\\ 

The first realization of supersymmetric hybrid inflation was proposed by Dvali et {\it al} \cite{key8} and  was based on the superpotential 
\begin{equation}
W = \kappa S(-\mu^2 + \overline{\phi}\phi)
\end{equation}
where $\overline{\phi}$, $\phi$ is a conjugate pair of superfields transforming as non-trivial representation of some gauge group $G$ under which the superfield $S$ is neutrally charged, the coupling parameter $\kappa$ and the mass scale $\mu$ can be taken to be real and positive . As was noted in \cite{key8} the superpotential Eq.(1) is of the most general form consistent with an $R-$symmetry under which $S\rightarrow e^{i\alpha}S$, $W\rightarrow e^{i\alpha}W$ and $\overline{\phi}\phi$ invariant.\\

The scalar potential derived from the superpotential Eq.(1) is 
\begin{equation}
V = \kappa^2\vert- \mu^2 + \overline{\phi}\phi\vert^2 + \kappa^2\vert S\vert^2(\vert\phi\vert^2 + \vert\overline{\phi}\vert^2) + D-terms
\end{equation}
Restricting ourselves to the $D-$flat direction and bringing $S$, $\phi$, $\overline{\phi}$ on the real axis : $S\equiv\sigma/\sqrt{2}$, $\phi = \overline{\phi} \equiv\chi/\sqrt{2}$ where $\sigma$, $\chi$ are normalized real scalar fields. The potential then takes the familiar form of the Linde's potential for hybrid inflation \cite{key7} but without the mass-term of $\sigma$ which is of crucial importance since it provides the slope of the valley of minima necessary for inflation .Hybrid inflation scenario with this superpotential has been extensively studied in the litterature (see \cite{key9} for a review). One way to obtain the valley of minima useful for inflation is to use the radiative corrections along the inflationary valley ($\chi = 0, \sigma > \sigma_c \equiv\sqrt{2}\mu$). The one loop effective potential is then given by :
\begin{equation}
V_{eff} = \kappa^2\mu^4\biggl[1 + \frac{\kappa^2}{16\pi^2}\biggl(ln\biggl(\frac{\kappa^2\sigma^2}{2\Lambda^2}\biggr) + \frac{3}{2} +\cdot\cdot\cdot\biggr)\biggr]
\end{equation}  

In this paper we propose an illustration of the above mentioned non-conventional cosmology by considering the model Eq.(3) in the context of the braneworld scenario. We study how the parameters of the model are then affected when the cosmological constraints from COBE are invoqued.\\

As was noticed above in the context of the braneworld picture with extra dimensions there may be a deviation of the effective four-dimensionl cosmology from the standard one. Indeed, if we assume that Einstein equations with a negative cosmological constant hold in $D$ dimensions and that matter fields are confined in the three-brane then the four-dimensional Einstein equation is given by  \cite{key4} 
\begin{equation}
G_{\mu\nu} = -\Lambda g_{\mu\nu} + \frac{8\pi}{M_p^2}T_{\mu\nu} + \biggl(\frac{8\pi}{M_5^3}\biggr)^2 S_{\mu\nu} - E_{\mu\nu} 
\end{equation}
where $T_{\mu\nu}$ is the energy momentum on the brane, $S_{\mu\nu}$ is a tensor that contains contributions that are quadratic in $T_{\mu\nu}$, and $E_{\mu\nu}$ corresponds to the projection of the five-dimensional Weyl tensor on the three-brane. The four-dimensional cosmological constant is related to the five-dimensional cosmological constant and the three-brane tension $\lambda$ as
\begin{equation}
\Lambda = \frac{4\pi}{M_5^3}\biggl(\Lambda_5 + \frac{4\pi}{3M_5^3}\lambda^2\biggr) 
\end{equation}
while the fundamental five-dimensional Planck scale and the effective four-dimensional one are related as
\begin{equation}
M_p = \sqrt{\frac{3}{4\pi}}\frac{M_5^3}{\sqrt{\lambda}} 
\end{equation}

In a cosmological context where our universe is a three brane, and the metric projected onto the brane is a homogeneous and isotropic, flat Robertson-Walker metric, the generalized Friedmann equation has the following form \cite{key5} 
\begin{equation}
H^2 = \frac{\Lambda}{3} + \biggl(\frac{8\pi}{3M_p^2}\biggr)\rho + \biggl(\frac{4\pi}{3M_5^3}\biggr)^2 \rho^2 + \frac{\epsilon}{a^4} 
\end{equation}
where $\epsilon$ is an integration constant arising from $E_{\mu\nu}$. During inflation the last term in the above equation will be neglected, moreover the cosmological constant is negligible in the early universe which means that $\Lambda_5\simeq -4\pi\lambda^2/3M_5^3$. The Friedmann equation takes then the final form
\begin{equation}
H^2 = \frac{8\pi}{3M_p^2}\rho\biggl[1 + \frac{\rho}{2\lambda}\biggr] 
\end{equation}

The new term in $\rho^2$ is dominant at high energies, but at low energies we must recover the standard cosmology, in particular during the nucleosynthesis, which leads to a lower bound on the value of the brane tension $\lambda\ge (1 MeV)^4$ . \\

In the slow-roll approximation  the generalized  expressions of the main quantities of the inflationary paradigm was derived in reference \cite{key10} in the case of the simplest chaotic inflationary scenario, and since the inflaton field $\sigma$ behaves during the inflationary phase of the hybrid model ($\chi = 0, \sigma > \sigma_c \equiv\sqrt{2}\mu)$ as in the chaotic inflation, these results will also be useful for oour analysis .\\

In the slow-roll approximation the slope and the curvature of the potential must satisfy the two constraints $\epsilon\ll 1$ and $\eta\ll 1$ where $\epsilon$ and $\eta$ are the two slow-roll parameters which are now defined by \cite{key10} 
\begin{eqnarray}
\epsilon & \equiv & \frac{M_p^2}{16\pi}\biggl(\frac{V'}{V}\biggr)^2\biggl[\frac{2\lambda(2\lambda + 2V)}{(2\lambda + V)^2}\biggr] \\
\eta & \equiv &  \frac{M_p^2}{8\pi}\biggl(\frac{V''}{V}\biggr)\biggl[\frac{2\lambda}{2\lambda + V}\biggr]  
\end{eqnarray}
 
The end of inflation will take place for a field value $\sigma_e$ such that
\begin{equation}
 \mbox{max}[\epsilon(\sigma_e) , \eta(\sigma_e)] = 1
\end{equation} 

However, in hybrid scenario inflation can end at $\sigma = \sigma_c$ if $\sigma_c \ge \sigma_e$. In the present case the two values are given by $\sigma_e = \bigl(M_p/(4\pi)^{3/2}\bigr)\bigl(\sqrt{\lambda}/\mu^2\bigr)$, $\sigma_c = \sqrt{2}\mu$. In what follows we will assume that inflation ends at $\sigma_c$, this is equivalent to the condition
\begin{equation}
\frac{\sqrt{\lambda}}{\mu^2} \le 0.6\times 10^2\frac{\mu}{M_p} 
\end{equation} 

The number of e-folds during inflation between two field values becomes \cite{key10} 

\begin{equation}
N \simeq -\frac{8\pi}{M_p^2}\int_{\sigma_1}^{\sigma_2} \frac{V}{V'}\biggl[1 + \frac{V}{2\lambda}\biggr]d\sigma 
\end{equation}

It is clear from Eqs.(8), (9), (10) and (13) that at low energies (i.e. $\lambda > V$) we retrieve the familiar expressions of the conventional cosmology. The main cosmological constraint comes from the amplitude of the scalar perturbations which is given in this new context by \cite{key10} 
\begin{equation}
A_s^2 \simeq \biggl(\frac{512\pi}{75M_p^6}\biggr)\frac{V^3}{V'^2}\biggl[\frac{2\lambda + V}{2\lambda}\biggr]^3\Biggl\vert_{k = aH} 
\end{equation}
where the right-hand side is evaluated at the horizon-crossing when the comoving scale equals the Hubble radius during inflation, which occurs approximately 55 e-folding before the end of inflation. In order to evaluate the corresponding inflaton field value $\sigma_*$ we take $N = 55$, $\sigma_2 = \sigma_c$ and $\sigma_1 = \sigma_*$ . Combining the result with Eq.(14) and using the COBE observed value $A_s = 2\times 10^{-5}$ we obtain 

\begin{equation}
\mu \simeq 4\times 10^{-4}M_p
\end{equation} 

Replacing this value in Eq.(12) yelds the uper bound on the brane tension

\begin{equation}
\lambda \lesssim (7\times 10^{14}GeV)^4
\end{equation}
which, in turn, implies an upper bound on the fundamental five-dimensional Planck scale

\begin{equation}
M_5 \lesssim 2\times 10^{-3}M_p
\end{equation}

Here we have hardly obtained a value of $M_5$ not smaller than $\mu$, as it should be in realistic models. This is indeed a consequence of the condition [Eq.(12)] from which we can deduce, using Eq.(6) that the two scales are related as $M_5\le {\cal O}(10) \mu$.\\

As was noticed above, in the early stages of the evolution of the universe, the extra term in the Friedmann equation [Eq.(8)] is dominant, which is equivalent to the condition $\kappa^2\mu^4 > \lambda$. Puting this condition together with the equation  (12) we get

\begin{equation}
\kappa \gtrsim 2\times 10^{-2}    
\end{equation} 

Although it is clear from Eq.(18) that no fine-tuning is required in this model, this seems not to be quite correct because of the usual cosmological constant constraint. However, this bound on the coupling parameter has been deduced by using the condition $\kappa^2\mu^4 >\lambda$ which is valid at high energy scales where it has been suggested that natural values of the cosmological constant $\Lambda$ are between $1$ TeV and $M_p$ (see e.g.\cite{key12} and references therein). This means that the corresponding vacuum energy density $\rho_{\Lambda}$ is between ($1$ TeV)$^4$ and $M_p^4$. Therfore, in this model, if we consider that $\rho_{\Lambda}=\kappa^2\mu^4$ as in the standard hybrid scenario (the mass parameter $\mu$ being given by Eq.(15)), the above value of $\kappa$ does not seem problematic.\\

The scale dependence of the perturbations is described by the spectral tilt

\begin{equation}
n_s - 1 \equiv \frac{dln A_s^2}{dln k} = 2\eta - 6\epsilon
\end{equation}
where the slow-roll parameters are given in Eqs.(9) and (10) . For this model Eq.(3) we obtain 
\begin{equation}
n_s = 0.98 \;\;\;  \mbox{for} \;\;\;\kappa \simeq 10^{-2}
\end{equation} 

Finally, from the ratio between the amplitude of tensor and scalar perturbations which is given by \cite{key11} 
\begin{equation}
\frac{A_t^2}{A_s^2} \simeq \frac{3M_p^2}{16\pi}\biggl(\frac{V'}{V}\biggr)^2\frac{2\lambda}{V} 
\end{equation}
It follows that the tensor perturbations are negligible . indeed, for the above values of the parameters we have
\begin{equation} 
A_t \simeq 2\times 10^{-4} A_s
\end{equation}

As a conclusion, we have obtained practically the same results as in the initial model \cite{key8}, which means that the advantages of such models can be preserved in the framework of the recently proposed braneword picture. 

To summarize we have reexamined a supersymmetric hybrid inflationary model in the context of the braneworld scenario. We found that the observational bounds are satisfied with an energy scale of inflation of order $10^{15}$ GeV. This can be achieved without any fine-tuning of the coupling parameter, provided that the fundamental five-dimensional Planck scale statisfies $M_5\le 2\times 10^{-3}M_p$ . We have also obtained an upper bound on the brane tension.

\end{document}